\documentclass[a4,14pt]{article}
\usepackage[margin=1in]{geometry}
\usepackage{subcaption}
\usepackage{amsmath}
\usepackage{hyperref}
\usepackage{amssymb}
\usepackage{braket}
\usepackage{upgreek}
\usepackage{graphicx}
\usepackage{float}
\usepackage{cite}
\usepackage{esvect}
\begin{document}
\pagenumbering{arabic}
\title{Mass Varying Neutrino Oscillation in Scalar-Gauss-Bonnet Gravity}
\author{H. Mohseni Sadjadi\footnote{mohsenisad@ut.ac.ir} and H. Yazdani Ahmadabadi\footnote{hossein.yazdani@ut.ac.ir}
	\\ {\small Department of Physics, University of Tehran,}
	\\ {\small P. O. B. 14395-547, Tehran 14399-55961, Iran}}
\maketitle
\begin{abstract}
We investigate how matter density affects neutrino oscillations by considering a mass-varying neutrino scenario
where the neutrino mass depends on a scalar field. This scalar field is non-minimally coupled to
the Gauss-Bonnet (GB) invariant, causing its profile to be implicitly influenced by the surrounding matter distribution.
Using data from solar neutrino experiments, we derive constraints on the model parameters, providing new insights
into the properties of mass-varying neutrino within the Gauss-Bonnet scalar-tensor framework.
\end{abstract}

\section{Introduction}\label{sec1}
The discovery of neutrino oscillations has established that neutrinos possess mass, with each flavor eigenstate
($\nu_e$, $\nu_\mu$, and $\nu_\tau$) forming a superposition of
mass eigenstates ($\nu_1$, $\nu_2$, $\nu_3$) \cite{Pontecorvo,Mikh-Smir,Wolf,Gonzalez-Garcia,Chakraborty,Capozziello}.
This three-flavor framework has been strongly supported by results from experiments such as the Sudbury Neutrino Observatory (SNO) \cite{SNO},
Super-Kamiokande (SK) \cite{Super-Kamiokande,Super-Kamiokande-Fukuda}, and KamLAND \cite{KamLAND}.

Despite its success, several anomalies—such as the $2\sigma$ difference in $\Delta m^2_{21}$ values measured from the solar and terrestrial experiments \cite{KamLAND,Smirnov-anom,Wurm}, and results observed in the LSND \cite{LSND} and MiniBooNE \cite{MiniBooNE1,MiniBooNE2} experiments—hint at possible physics beyond the standard oscillation paradigm.
These discrepancies have spurred interest in extensions involving non-standard neutrino interactions (NSIs) \cite{Miranda,Dev}.
Some of the aforementioned anomalies could potentially arise from varying neutrino mass observed in different environments with various densities.

A well-established modification of flavor evolution in matter is the Mikheyev–Smirnov–Wolfenstein (MSW) effect
\cite{Mikh-Smir,Wolf-MSW,Mikh-Smir-MSW},
where forward scattering of neutrinos off electrons induces an effective potential that alters the flavor conversion probability.
Since this effect arises from weak interactions, it is natural to ask whether other types of interactions—particularly
those mediated by scalar fields—might induce similar or competing effects.

Scalar fields coupled to neutrinos have been studied in the context of mass-varying neutrino (MaVaN) models \cite{Stephenson,Halprin},
often motivated by the intriguing proximity between the dark energy scale and neutrino masses.
These models propose that neutrino masses evolve dynamically due to their interaction with a cosmological scalar field,
potentially linking cosmic acceleration to neutrino physics \cite{Kaplan,Barger-DE,Fardon,MohseniSadjadi,MohseniSadjadi1,MohseniSadjadi2,YA,Mazia}.
This connection also suggests that varying neutrino masses could play a significant role in cosmological evolution \cite{Hung,Brookfield,Geng}.
In such scenarios, the scalar field can be sensitive to the local matter density, thereby inducing environment-dependent
neutrino masses and modifying the oscillation behavior in a manner reminiscent of the MSW effect \cite{Sadjadi-Khosravi}.

Building on this idea, some scalar-tensor theories of gravity—such as chameleon\cite{ch} and symmetron\cite{sym} models—have been considered,
where scalar fields with screening mechanisms mediate interactions between neutrinos and the gravitational sector
\cite{MohseniSadjadi-Yazdani1,MohseniSadjadi-Yazdani2,MohseniSadjadi-Yazdani3,MohseniSadjadi-Yazdani4}.
Chameleon and symmetron models, in which matter density couples to the scalar field,
enable the scalar field to remain consistent with local gravity constraints while still influencing
neutrino physics in low-density environments.

In this work, we propose a novel mechanism for mass-varying neutrinos oscillations within the framework of a well-known scalar–tensor model, i.e., scalar–Gauss–Bonnet (sGB) gravity. The sGB model extends general relativity by coupling a scalar field to the Gauss–Bonnet (GB)
invariant. Although the GB term (a total divergence in 4D) does not alter Einstein’s equations alone,
 its coupling to a scalar field modifies the equations of motion, producing geometry-dependent scalar field profile.
 Such couplings have recently been employed in black hole and neutron star scalarization scenarios
\cite{GB,GB1,GB2,GB-Minamitsuji,GB-Peng,GB-MohseniSadjadi}. Unlike conformally coupled models (e.g., chameleon/symmetron),
where matter density appears explicitly in the equation of motion,
in the sGB model it appears implicitly via curvature terms through the GB term. This results in
an environment-dependent mass for the neutrinos. We show that this mechanism
suppresses mixing in high-density regions while flavor oscillations occur in regions of lower density.
In conclusion, by analyzing solar neutrino experimental data, we establish constraints on the model parameters, yielding novel insights into mass-varying neutrino flavor conversion within the GB scalar-tensor formalism.
The derived constraints agree well with the LMA-MSW solution for the solar neutrino problem, confirming that the sGB model can reconcile observations without introducing significant deviations from the standard oscillation predictions.

The paper is organized as follows. In Section \ref{sec2}, we derive the scalar field
profile both inside and outside the spherically symmetric matter distribution like the Sun.
In Section \ref{sec3}, we analyze neutrino flavor conversion in the presence of the scalar field. In Section \ref{sec4}, we present numerical results and discuss constraints from solar neutrino experiments.
Finally, in Section \ref{sec5}, we summarize our results.

Throughout this paper, we work in natural units where $\hbar = c = 1$.

\section{Scalar Field Behavior}\label{sec2}
We consider a scalar field $\phi$ that interacts with non-relativistic matter through the coupling to the GB invariant consisting of quadratic terms of the Ricci scalar, Ricci tensor, and Riemann tensor. Additionally, the scalar field couples to neutrinos via the neutrino mass term. The action for this system is given by
\begin{eqnarray}\label{eqn1}
S = \int d^4x \sqrt{-g} \left[\frac{M_p ^2 \mathcal{R}}{2} - \frac{1}{2} g^{\mu\nu} \partial_\mu \phi \partial_\nu \phi -
f(\phi)\mathcal{G} + \kappa_i \phi^2 \bar{\nu}_i \nu_i - i \bar{\nu}_i \gamma^\mu D_\mu \nu_i + \mathcal{L} \left(g_{\mu\nu} , \Psi_i\right)\right],
\end{eqnarray}
where $\mathcal{G} = \mathcal{R}^2 - 4\mathcal{R}^{\mu\nu} \mathcal{R}_{\mu\nu} + \mathcal{R}^{\mu\nu\rho\sigma} \mathcal{R}_{\mu\nu\rho\sigma}$
is the GB invariant, and $\kappa_i$'s are the coupling coefficients (with dimensions of $\text{eV}^{-1}$) between the scalar field and
the $i$th neutrino eigenstate.
These coefficients must be chosen to comply with the neutrino mass reported by neutrino oscillation experiments.
The Lagrangian $\mathcal{L}(g_{\mu\nu},\Psi_i)$ encompasses all matter components.
The non-trivial coupling between the scalar field and the GB invariant is governed by the function $f(\phi)$, which might have a quadratic dependence to the scalar field which mimics an effective mass term
\begin{eqnarray}\label{eqn2}
f(\phi)= \frac{1}{2} \xi \phi^2,
\end{eqnarray}
where $\xi$ is a parameter with units of [length]$^2$.

We have considered a non-minimal coupling between massive neutrinos and the scalar field, which results in the neutrino mass
 acquiring a quadratic dependence on the scalar field (from Eq. (\ref{eqn1})).
While the coupling between the scalar field and matter fields is primarily phenomenological, there are theoretical frameworks
that propose more fundamental origins for such interactions.
For instance, in the context of the seesaw mechanism for neutrino masses \cite{Fardon,Bi}, the Lagrangian for neutrinos can be expressed as
\begin{eqnarray}\label{eqn3}
\mathcal{L}_{\text{neutrino}} = h_{ij} \bar{l}_{Li} \nu_{Rj} H + \frac{1}{2} M_{ij} (\phi) \bar{\nu}^c_{Ri} \nu_{Rj} + H.c.,
\end{eqnarray}
where $\nu_R$ is the right-handed sterile neutrino.
In such models, the neutrino masses are functions of an exotic field, with $M_{ij}(\phi) = M_{ij}/A(\phi)$.
The Dirac mass of neutrinos is given by $h_{ij} < H >$.
In addition, $l_L$ and $H$ denote the left-handed lepton and Higgs-doublets.
By integrating the sterile neutrinos, a $\phi$-dependent mass term, proportional to the function $A(\phi)$,
is generated for the active neutrinos through a dimension-5 operator \cite{Bi}.
In our current model, the neutrino mass term is proportional to $\phi^2$.
As discussed in Ref. \cite{Cleaver}, such a mass term can naturally
emerge in effective field theory (EFT) from higher-dimensional operators in seesaw and supergravity scenarios.
Additionally, a scalar field that is massless at tree level may acquire mass through loop corrections.
For example, radiative corrections from gauge fields can induce masses for the scalar degrees of freedom.
This mechanism imposes upper bounds on the scale of new physics related to neutrinos \cite{Fardon,Vissani,Maki}.
However, the specifics of these bounds vary depending on the model and lie outside the scope of this paper.

By varying the action (\ref{eqn1}) with respect to the scalar field $\phi$, we obtain the equation of motion
\begin{eqnarray}\label{eqn4}
\square \phi - \mathcal{G} f_{,\phi} + 2\kappa_i \bar{\nu}_i \nu_i \phi=0.
\end{eqnarray}
To begin our analysis, we first consider the Schwarzschild metric, which describes a general static, spherically symmetric object such as the Sun.
Detailed calculations are provided in the remainder of this section, where we summarize the key results and offer physical
 interpretations by presenting and analyzing relevant figures.

Astrophysical objects are generally  spherical in shape.
Therefore, we adopt the metric for a non-relativistic, static, and spherically symmetric object with mass $M_\odot$ and radius $R_\odot$.

The interior metric of such a spherically symmetric body is given by
\begin{eqnarray}\label{eqn5}
ds^2_{\text{in}} = -\frac{1}{4} \left(3\sqrt{1- \frac{r_s}{R_\odot}} - \sqrt{1- \frac{r^2 r_s}{R_\odot^3}}\right)^2 dt^2 +
\left(1- \frac{r^2 r_s}{R_\odot^3}\right)^{-1} dr^2 + r^2 d\theta^2 + r^2 \sin^2 \theta d\varphi^2,
\end{eqnarray}
where $r_s = 2GM_\odot$ represents the Schwarzschild radius of the object.
Outside the body, however, we assume that the density is significantly lower than the interior density.
In this region, the metric can be described by the Schwarzschild exterior solution
\begin{eqnarray}\label{eqn6}
ds^2_{\text{out}} = - \left(1- \frac{r_s}{r}\right) dt^2  + \left(1- \frac{r_s}{r}\right)^{-1} dr^2 + r^2 d\theta^2 +
 r^2 \sin^2 \theta d\varphi^2.
\end{eqnarray}
To analyze the scalar field equation (\ref{eqn4}), we compute the GB invariant for both the interior and exterior regions.
Inside the object, the GB invariant is given by
\begin{eqnarray}\label{eqn7}
\mathcal{G}_{\text{in}}(r) = \frac{24 r_s^2 \left[R_\odot^3 \left(-3 \sqrt{1-\frac{r_s}{R_\odot}} \sqrt{1-\frac{r^2 r_s}{R_\odot^3}}
 - 1\right) + r^2 r_s\right]}{r^2 R_\odot^6 r_s + 8 R_\odot^9 - 9 R_\odot^8 r_s},
\end{eqnarray}
while outside the object, it takes the form
\begin{eqnarray}\label{eqn8}
\mathcal{G}_{\text{out}} (r) = \frac{12 r_s^2}{r^6}.
\end{eqnarray}
By neglecting the term proportional to the neutrino mass in Eq. (\ref{eqn4}) and substituting the GB invariant for each case,
the scalar field equation of motion \textbf{inside} the spherical object becomes
\begin{eqnarray}\label{eqn9}
\begin{split}
& r^\prime \left(1 - r_s^\prime r^{\prime 2}\right) \left(r_s^\prime r^{\prime 2} - 9 r_s^\prime + 8\right) \frac{d^2\hat{\phi}}{dr^{\prime 2}}
 \\&- \left[4 r_s^{\prime 2} r^{\prime 4} - 27 r_s^{\prime 2} r^{\prime 2} + 18 r_s^\prime - 16 - 3 r_s^\prime r^{\prime 2}
  \left(\sqrt{1 -  r_s^\prime} \sqrt{1 - r_s^\prime r^{\prime 2}} - 7\right)\right] \frac{d\hat{\phi}}{dr^\prime} \\&  -
  24 r_s^{\prime 2}  r^\prime \left[r_s^\prime r^{\prime 2} -3 \sqrt{1 -  r_s^\prime} \sqrt{1 - r_s^\prime r^{\prime 2}} - 1 \right]
    \xi^\prime\hat{\phi} = 0,
\end{split}
\end{eqnarray}
where we have redefined dimensionless parameters
\begin{eqnarray}\label{eqn10}
\hat{\phi}(r) = \frac{\phi(r)}{\phi_{\text{in}}(0)} ,~~~~~~~~ r^\prime = \frac{r}{R_\odot} ,~~~~~~~~  r_s^\prime = \frac{r_s}{R_\odot},~~~~~~~~  \xi^\prime = \frac{\xi}{R_\odot^2} .
\end{eqnarray}
The function $f(\phi)$ is in principle arbitrary, but with the above scalar field normalization, the function $f(\hat{\phi}) = \frac{1}{2} \xi^\prime \hat{\phi}^2$ remains a dimensionless function of a dimensionless argument.

Similarly, the scalar field equation of motion \textbf{outside} the sphere simplifies to
\begin{eqnarray}\label{eqn11}
	\left(1- \frac{r_s^\prime}{r^\prime}\right) \frac{d^2 \hat{\phi}}{dr^{\prime 2}} +
 \frac{1}{r^\prime} \left(2 - \frac{r_s^\prime}{r^\prime}\right) \frac{d\hat{\phi}}{dr^\prime}  - \frac{12 r_s^{\prime 2}}{r^{\prime 6}} \xi^\prime \hat{\phi} = 0.
\end{eqnarray}

To solve equations (\ref{eqn9}) and (\ref{eqn11}), we need to establish appropriate boundary conditions.
These conditions ensure that the solution remains non-singular at the origin ($d\hat{\phi}_{\text{in}}/dr^\prime \to 0$ at $r^\prime =0$) and that the force on a test particle vanishes at infinity ($\hat{\phi}_{\text{out}} \to \hat{\phi}_{\infty} = const.$ as $r^\prime \to \infty$).
Assuming $r^\prime_s \ll 1$, the general regular solution of the scalar field inside the sphere is approximately given by
\begin{eqnarray}\label{eqn12}
\hat{\phi}_{\text{in}}(r^\prime) = \frac{\sinh \left(2\sqrt{3 \xi^\prime} r_s^{\prime} r^\prime\right)}{2\sqrt{3 \xi^\prime}
 r_s^{\prime} r^\prime}.
\end{eqnarray}
With a similar procedure, the solution for the scalar field outside the object is given by
\begin{eqnarray}\label{eqn13}
\hat{\phi}_{\text{out}}(r^\prime) = A \frac{J_{-\frac{1}{4}} \left[\frac{\sqrt{3\xi^\prime} r^\prime_s}{r^{\prime 2}}\right]}
{\sqrt{r^\prime}}  +  B\frac{J_{\frac{1}{4}} \left[\frac{\sqrt{3\xi^\prime} r^\prime_s}{r^{\prime 2}}\right]}{\sqrt{r^\prime}},
\end{eqnarray}
where $J_n(z)$ is the Bessel function of the first kind.
With this solution, the boundary condition is satisfied at infinity, i.e., $\hat{\phi} (r^\prime \to \infty) \to \hat{\phi}_{\infty}= const.$ .
The constant value is explicitly given by $\hat{\phi}_{\infty} = \frac{A}{\Gamma[3/4]} \left[\frac{2}{\sqrt{3\xi^\prime}r_s^\prime}\right]^{\frac{1}{4}}$.
The value of $\hat{\phi}_\infty$ becomes negligible simply as $A \to 0$, which is relevant particularly to the possible role of the scalar field in the context of scalarization, studied within GB gravity.

The boundary conditions at the sphere's surface ($r^\prime = 1$), which require the continuity of the field $\hat{\phi}$ and its first derivative, determine the integration constants $A$ and $B$.
These constants are evaluated as
\begin{eqnarray}\label{eqn14}
A = \frac{\pi}{\sqrt{8}} \left(\cosh \left[2 \sqrt{3 \xi^\prime} r_s^\prime\right]
 J_{\frac{1}{4}}\left[\sqrt{3 \xi^\prime} r_s^\prime\right] - \sinh \left[2\sqrt{3 \xi^\prime} r_s^\prime\right]
 J_{\frac{5}{4}}\left[\sqrt{3 \xi^\prime} r_s^\prime\right]\right),
\end{eqnarray}
and
\begin{eqnarray}\label{eqn15}
B = -\frac{\pi}{\sqrt{8}} \left(\cosh \left[2 \sqrt{3 \xi^\prime} r_s^\prime\right]
 J_{-\frac{1}{4}}\left[\sqrt{3 \xi^\prime} r_s^\prime\right] + \sinh \left[2\sqrt{3 \xi^\prime} r_s^\prime\right] J_{-\frac{5}{4}}\left[\sqrt{3 \xi^\prime} r_s^\prime\right]\right).
\end{eqnarray}
These expressions for $A$ and $B$ directly yield the value of the field at the boundary, confirming the first continuity condition $\hat{\phi}_{\text{in}} (1) = \hat{\phi}_{\text{out}}(1) = \frac{\sinh \left(2\sqrt{3 \xi^\prime} r_s^{\prime}\right)}{2\sqrt{3 \xi^\prime} r_s^{\prime}}$.
Furthermore, these constants ensure the continuity of the field gradient, satisfying the second boundary condition $\frac{d\hat{\phi}_{\text{in}}}{dr^\prime}|_{r^\prime=1} = \frac{d\hat{\phi}_{\text{out}}}{dr^\prime}|_{r^\prime=1} = \cosh(2\sqrt{3 \xi^\prime} r_s^{\prime}) - \frac{\sinh \left(2\sqrt{3 \xi^\prime} r_s^{\prime}\right)}{2\sqrt{3 \xi^\prime} r_s^{\prime}}.$

It is worth noting that the normalization (\ref{eqn10}) for the scalar field seems vital because the sGB theory does not fix the field's value at the center of the sphere (i.e., at $r' = 0$), leaving $\phi_{\text{in}}(0)$ as a parameter that can be absorbed into others.
Consequently, the dimensionless field $\hat{\phi}(r')$ represents the scalar field behavior relative to this central value $\phi_{\text{in}}(0)$.
The large values of $\hat{\phi}(r^\prime)$ outside the Sun do not necessarily imply large physical values of the scalar field $\phi(r)$ itself; they indicate that the field profile grows relative to its central boundary value.

Although normalizing with respect to the reduced Planck mass $M_p$ is mathematically possible, it would leave the normalized field $\hat{\phi}(r)$ unchanged, since it is defined as the ratio of $\phi(r)$ to $\phi_{\text{in}}(0)$—a quantity invariant under rescaling by any fundamental constant.
Nevertheless, the current normalization is advantageous, as it intrinsically accounts for the theory’s undetermined boundary value $\phi_{\text{in}}(0)$, offering a more natural and intuitive depiction of the field’s spatial variation in relation to its central value.

Moreover, to ensure the scalar field's backreaction on the spacetime geometry is negligible, the conditions $(\partial_r \phi)^2 \ll M_p^2 \mathcal{R}$ and $\xi \phi^2 \mathcal{G} \ll M_p^2 \mathcal{R}$ must be satisfied.
Expressed via some of dimensionless model parameters (\ref{eqn10}), these constraints become
\begin{eqnarray}\label{eqn16}
\begin{split}
&\frac{\phi^2_{\text{in}}(0)}{M_p^2} \ll \frac{R_\odot^2 \mathcal{R}}{(d\hat{\phi}/dr^\prime)^2}, \\& \frac{\phi^2_{\text{in}}(0)}{M_p^2} \ll \frac{\mathcal{R}}{\mathcal{G} \xi^\prime R^2_\odot \hat{\phi}^2}.
\end{split}
\end{eqnarray}
These final forms provide direct upper bounds on the central field value, $\phi_{\text{in}}(0)$, relative to the reduced Planck mass, $M_p$.
The bounds depend on the solar radius squared, $R_\odot^2$, and the local geometric terms, namely the Ricci curvature $\mathcal{R}$ and the GB invariant $\mathcal{G}$.

\begin{figure}[H]
	\begin{subfigure}{.5\textwidth}
		\centering
		\includegraphics[scale=0.46]{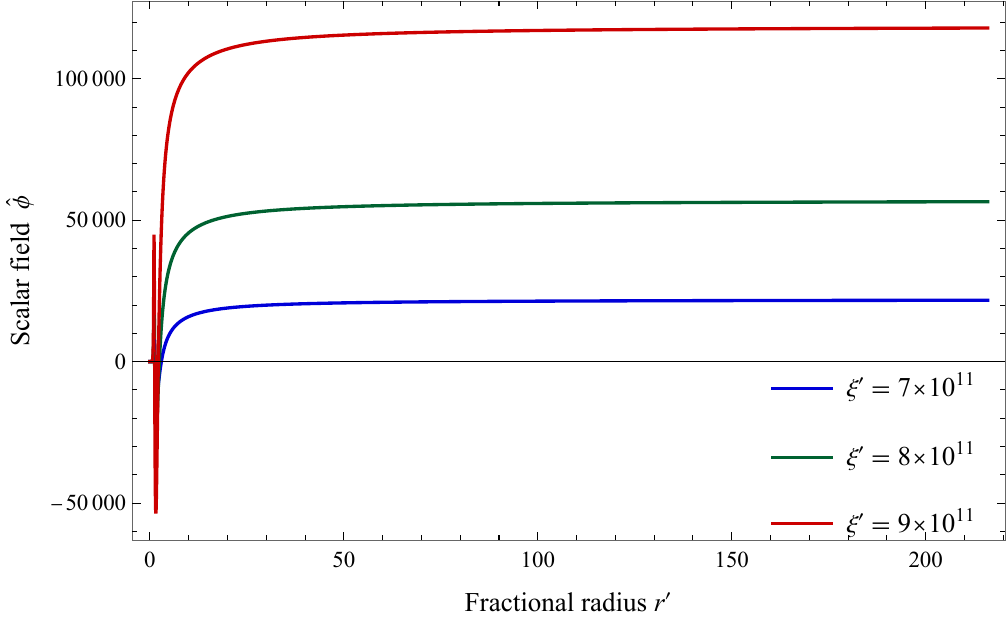}
		\label{fig1-a}
		\caption{\footnotesize{}}
	\end{subfigure}
	\begin{subfigure}{.5\textwidth}
		\centering
		\includegraphics[scale=0.51]{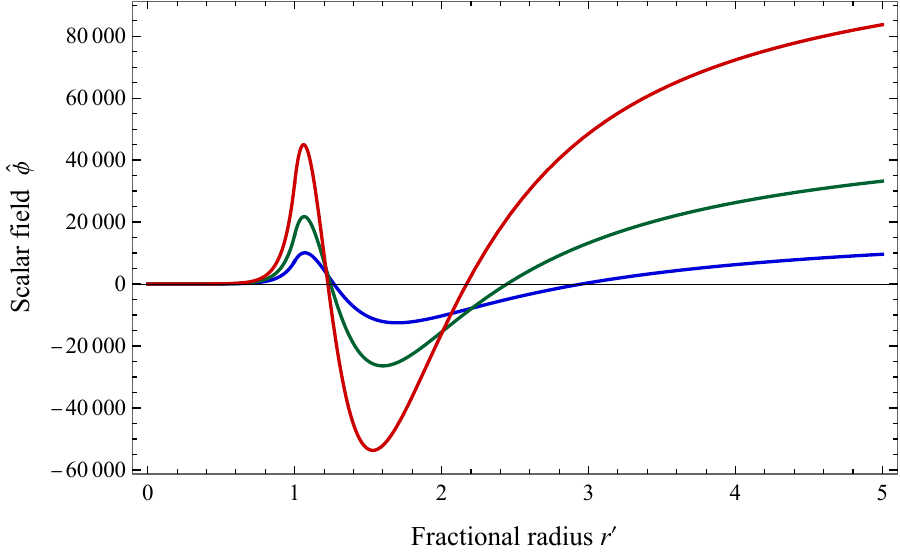}
		\label{fig1-b}
		\caption{\footnotesize{}}
	\end{subfigure}
	\caption{\footnotesize{The dimensionless scalar field profile as a function of the dimensionless fractional radius $r^\prime$.
			These cases correspond to $ r_s^{\prime}= 4.17\times 10^{-6}$ (the solar case) and $\xi^\prime \in \{7 \times 10^{11}, 8 \times 10^{11}, 9\times 10^{11}\}$.
			At large distances, the scalar field exhibits smooth behavior, see panel (a).
			Panel (b) is drawn to illuminate the dependence of $\hat{\phi}(r^\prime)$ on the parameters inside and near the surface of the object.}}
	\label{fig1}
\end{figure}

A numerical approach to solving the scalar field differential equation allows us to explore the field's behavior for the model parameter and make predictions about the neutrino characteristics, such as the neutrino mass and the neutrino mass-squared splitting inside and around the Sun.
With numerical computations, we plot Fig. \ref{fig1}, which shows the scalar field as a function of the fractional radius $r^\prime$ for three possible values of the parameter $\xi^\prime$.
For all values of $\xi^\prime$, the scalar field well within the object becomes $\hat{\phi}_{\text{in}} \sim 1$, which is negligibly small compared to the field values outside, whereas $\hat{\phi}$ exhibits smooth behavior at infinity.
As $\xi^\prime$ increases, the asymptotic absolute value tends to the higher amounts.

Now, as illustrated in Fig. \ref{fig1} and governed by Eq. (\ref{eqn16}), a numerical constraint can be placed on the central scalar field value, $\phi_{\text{in}}(0)$.
The analysis employs the field solutions from Eqs. (\ref{eqn12})-(\ref{eqn15}), which yield the ranges $\hat{\phi}_{\text{in}} \sim 1 - 10^4$ and $d\hat{\phi}_{\text{in}}/dr^\prime \sim 0 - 10^5$ from the solar center to its surface.
In conjunction with the characteristic parameters $\xi^\prime \sim 10^{11} - 10^{12}$, $\mathcal{R}_{\text{in}} \sim 10^{-36} - 10^{-34}~\text{eV}^2$, and $\mathcal{G}_{\text{in}} \sim 10^{-72} - 10^{-68}~\text{eV}^4$ and $R_\odot\simeq 8.6\times 10^{42} M_p^{-1}$ for the Sun, (\ref{eqn16}) constrains the field to satisfy $\phi_{\text{in}}(0) \ll 10^{-8} M_p$. This ensures a negligible backreaction on the metric.
We will use these results to study mass-varying neutrino oscillation in the background of astrophysical objects like the Sun.
Note that although the Sun is weakly gravitating, a Newtonian or post-Newtonian treatment may simplify the analysis.
Still, we employed the Schwarzschild solution for consistency with the curvature computation because the GB invariant involves higher curvature terms.
This approach was chosen to maintain a smooth transition across the boundary.

\section{Matter Effects on Flavor Change}\label{sec3}
Neutrinos serve as a fascinating example of mixed particles, where the flavor eigenstates—participating in weak interaction vertices—are not the same as the mass eigenstates that govern their propagation.
This distinction arises because neutrinos propagate as mass eigenstates, each acquiring a phase that depends on their respective masses.
The phenomenon of neutrino mixing connects these two sets of states through a unitary transformation.
Mathematically, the mixing is expressed as
\begin{eqnarray}\label{eqn17}
\ket{\nu_\alpha (r)} = \sum_i U_{\alpha i} \ket{\nu_i (r)},
\end{eqnarray}
with coefficients given by the elements $U_{\alpha i}$ of the Pontecorvo-Maki-Nakagawa-Sakata (PMNS) matrix.
This matrix includes the mixing angles and phases that describe the relationship between the flavor and mass eigenstates.
The mixing mechanism is responsible for the phenomenon of neutrino oscillations, where neutrinos change their flavor as they propagate through space.
When a neutrino is produced in a weak interaction process with a specific flavor $\alpha$ at the source, it propagates as a superposition of mass eigenstates.
The dynamics of a neutrino state $\ket{\nu_i(r)}$ in a background gravitational field and matter potential are governed by a Schr$\ddot{\text{o}}$dinger-like equation
\begin{eqnarray}\label{eqn18}
i \frac{d}{dr} \ket{\nu_i(r)} = \mathcal{H}_{\text{eff.}} \ket{\nu_i(r)},
\end{eqnarray}
where the effective Hamiltonian is considered the sum of various modified contributions $\mathcal{H}_{\text{eff.}} = \mathcal{H}_{\text{vac.}} + \mathcal{H}_{\text{mat.}}$.
The following subsections investigate the mechanisms of flavor oscillation in vacuum and flavor change in matter separately.

\paragraph{Modified vacuum case}
In many experimental scenarios, a simplified two-flavor approximation is sufficient to describe neutrino oscillations with a good accuracy.
Under this assumption, the probability of flavor oscillation simplifies to
\begin{eqnarray}\label{eqn19}
P_{\alpha \beta} = \sin^2 (2\theta) \sin^2 \left(\frac{\Phi_{21}}{2}\right).
\end{eqnarray}
We begin with the phase difference for neutrino oscillations in curved spacetime, derived from the covariant Dirac equation and originally presented in our previous works \cite{MohseniSadjadi-Yazdani1,MohseniSadjadi-Yazdani2,MohseniSadjadi-Yazdani3,MohseniSadjadi-Yazdani4}.
The covariant form of the phase difference is given by
\begin{eqnarray}\label{eqn20}
\begin{split}
&\Phi_{ij} = \int^{x_B}_{x_A} p_\mu dx^\mu \\& ~~~~ = \int^{r_B}_{r_A} \frac{\Delta m_{ij}^2(r)}{2E_\nu} dr.
\end{split}
\end{eqnarray}
For a system with two neutrino flavors, the influence of sGB gravity alters the vacuum Hamiltonian.
This modified Hamiltonian, when transformed into the flavor basis using the PMNS matrix $U$, is given by the equation
\begin{eqnarray}\label{eqn21}
\mathcal{H}_{\text{vac.}} = \frac{\Delta m_{21}^2(r)}{4E_\nu}
\begin{pmatrix}
-\cos 2\theta & \sin 2\theta \\
\sin 2\theta & \cos 2\theta
\end{pmatrix}.
\end{eqnarray}
To model the scalar field's effects, we introduce a position-dependent mass term: $\Delta m_{21}^2(r) = \Delta\kappa_{21}^{\prime 2} \hat{\phi}^4(r)$ (based on the neutrino mass term in Eq.(\ref{eqn1})).
This term employs a modified coupling parameter, $\kappa_{i}^{\prime} \equiv \kappa_i \phi^2_{\text{in}}(0)$ (in eV), which encapsulates the central field value $\phi_{\text{in}}(0)$.
Because this value is unconstrained by the sGB model, absolute neutrino masses remain indeterminate without additional assumptions.
Therefore, our fitting procedure against observational data—a primary goal of this paper—directly yields constraints on the parameter $\Delta \kappa_{21}^{\prime 2}$.

Unlike in flat spacetime, where $E_\nu$ is the energy for any observer, the phase in the presence of gravity must be expressed using the local energy $E_l(r_B)$ measured at the detector (placed at $r_B$).
This local energy, defined as $E_l(r_B) = E_\nu / \sqrt{g_{tt}(r_B)} = E_\nu / \sqrt{1 - r_s/r_B}$, incorporates gravitational redshift to ensure consistency with the spacetime metric.
Using the weak-field approximation $1 / \sqrt{1 - r_s / r_B} \approx 1 + r_s / 2r_B$, we derive the following result for the phase difference (\ref{eqn20})
\begin{eqnarray}\label{eqn22}
\begin{split}
&\Phi_{21} = \int^{r_B}_{r_A} \mathcal{H}_{\text{vac.}} dr\\&~~~~ = \frac{R_\odot \Delta \kappa_{21}^{\prime 2}}{2E_l(r^\prime_{B})} \left[1 + \frac{r^\prime_s}{2r^\prime_B}\right] \int_{r^\prime_A}^{r^\prime_B} \hat{\phi}^4 (\text{r}^\prime) d\text{r}^\prime.
\end{split}
\end{eqnarray}
This expression employs the definition of the dimensionless fractional radii, previously defined by Eq.(\ref{eqn10}).

\paragraph{Matter effects added}
In matter, the Dirac equation governing neutrino propagation becomes non-diagonal in the flavor basis.
This crucial departure from the vacuum case stems from coherent forward scattering of neutrinos off particles in matter, a phenomenon that can induce resonant flavor oscillations, as famously described by the Mikheyev-Smirnov-Wolfenstein (MSW) effect \cite{Wolf-MSW,Mikh-Smir-MSW}.
Consequently, the equation of motion for a two-flavor neutrino system ($\nu_e, \nu_\mu$) traveling through matter must be modified to include a matter potential. This results in the following Schr$\ddot{\text{o}}$dinger-like evolution equation
\begin{eqnarray}\label{eqn23}
\begin{split}
& i \frac{d}{R_\odot dr^\prime}
\begin{pmatrix}
		\nu_e \\
		\nu_\mu
\end{pmatrix}
	\\& = \frac{1}{2}
	\left[\frac{\Delta \kappa_{21}^{\prime 2} \hat{\phi}^4(r^\prime_B)}{2E_l(r_B^\prime)} \left[1 + \frac{r^\prime_s}{2r^\prime_B}\right]
\begin{pmatrix}
		-\cos 2\theta&\sin 2\theta\\
		\sin 2\theta& \cos 2\theta\\
\end{pmatrix}
	+  \sqrt{2} G_F N_e(r^\prime)
\begin{pmatrix}
		1&0\\
		0&-1\\
\end{pmatrix}
	\right]
\begin{pmatrix}
		\nu_e\\
		\nu_\mu
\end{pmatrix}
	.
\end{split}
\end{eqnarray}
The first term in the Hamiltonian represents the vacuum oscillations, now incorporating corrections from the sGB gravity.
The second term is the matter potential, which originates from the weak interaction charge-current contribution; it affects electron-neutrinos ($\nu_e$) differently than muon- or tau-neutrinos ($\nu_\mu$ or $\nu_\tau$) because only $\nu_e$ can undergo charged-current scattering on electrons.
Here, $N_e$ is the electron number density, and $G_F \simeq 1.17 \times 10^{-5} ~\text{GeV}^{-2}$ is the Fermi constant.

By consolidating these terms, we can express the effective Hamiltonian in a more compact form
\begin{eqnarray}\label{eqn24}
\mathcal{H}_{\text{eff.}} = \frac{\Delta \kappa_{21}^{\prime 2} \hat{\phi}^4(r^\prime_B)}{4E_l(r^\prime_B)} \left[1 + \frac{r^\prime_s}{2r^\prime_B}\right]
\begin{pmatrix}
-\cos 2\theta + A & \sin 2\theta \\
\sin 2 \theta & \cos 2\theta - A
\end{pmatrix}
,
\end{eqnarray}
where the dimensionless quantity
\begin{eqnarray}\label{eqn25}
A = \pm \frac{2\sqrt{2} G_F N_e E_l(r^\prime_B)}{\Delta \kappa_{21}^{\prime 2} \hat{\phi}^4(r^\prime_B)}\left[1 - \frac{r^\prime_s}{2r^\prime_B}\right]
\end{eqnarray}
parameterizes the strength of the matter interaction relative to the vacuum oscillation term.
The positive sign ($+$) corresponds to neutrinos, while the negative sign ($-$) corresponds to antineutrinos, reflecting the fact that matter affects particles and antiparticles differently. It is important to note that any term proportional to the identity matrix in the Hamiltonian contributes only an overall, unobservable phase to the wavefunction and is therefore omitted from this physical expression without affecting the oscillation probabilities.

To verify our generalized derivation, we confirm that it reduces to the standard MSW effect under the appropriate limits: flat spacetime ($r^\prime_s \to 0$) and a limiting constant scalar field on detector ($\hat{\phi}(r_B^\prime) \to \hat{\phi}_{\infty}$).
These conditions cause the gravitational factor to be suppressed and the mass term to revert to the standard $\Delta m_{21}^2$, respectively.
Consequently, the effective Hamiltonian simplifies to its well-known form.
This agreement serves as a crucial check on the correctness of our model, which includes additional gravitational and scalar interactions.

In regions of high electron density, such as the interior of a star, the scalar field $\hat{\phi}_{\text{in}} \to 1$, which is negligibly small compared to the other values (see Fig. \ref{fig1}), causing the parameter $A$ (Eq. \ref{eqn25}) to dominate in the Hamiltonian.
In this limit, the diagonal elements become more significant, leading to mixing suppression.
This means that neutrino oscillations are less pronounced in dense matter regions.
Conversely, in regions of lower electron density, such as outside the sphere, the scalar field $\hat{\phi}$ is activated and takes on larger values.
In this regime, the parameter $A$ becomes negligible, and the Hamiltonian simplifies to the ordinary flavor oscillation case.
In a case when $A\simeq\cos2\theta$, the diagonal elements are zero and a resonant enhanced oscillation is resulted; hence, the maximally mixing is expected.

We turn to the problem of neutrino flavor conversion in matter, incorporating the effects of both matter and sGB gravity.
In the standard approach, matter effects are accounted for by diagonalizing the Hamiltonian, which yields the effective mass-squared difference and the effective mixing angle.
These quantities are given by
\begin{eqnarray}\label{eqn26}
\tan 2\theta_{M} = \frac{\sin 2\theta}{\cos 2\theta - A},
\end{eqnarray}
and
\begin{eqnarray}\label{eqn27}
\Delta m^2_M = \left[\Delta \kappa_{21}^{\prime 2} \hat{\phi}^4(r^\prime_B)\right] \sqrt{ 1 + A^2 - 2A\cos 2\theta}.
\end{eqnarray}
Here, $\theta_{M}$ represents the effective mixing angle in matter, and $\Delta m^2_{M}$ is the effective mass-squared difference, both of which depend on the matter potential parameter $A$ and the vacuum mixing angle $\theta$.
The MSW effect becomes significant for neutrino energies higher than the resonance energy, i.e., $E_\nu > E_{\text{res.}}(r^\prime)$.
In this regime, the electron-neutrino survival probability $P_{ee}(E_\nu)$ is well-defined and can be calculated using the effective parameters.
To investigate neutrino flavor conversion in the presence of the non-standard couplings of neutrinos, we follow the methodology outlined in Ref. \cite{MohseniSadjadi-Yazdani3}.
This approach leads to the following expression for the survival probability
\begin{eqnarray}\label{eqn28}
P_{ee}(E_\nu) = \frac{1}{2} \left[1 + \cos 2\theta_{M} \cos 2\theta\right].
\end{eqnarray}
At lower energies, neutrinos oscillate in the vacuum regime, where matter effects are negligible.
In this case, the survival probability $P_{ee}(E_\nu)$ exhibits a smooth transition between the vacuum and matter-dominated regimes.
For solar neutrinos, the energy ranges of $pp$, $^7$Be, and $pep$ neutrinos lie below the resonance region, meaning that matter effects are relatively small for these components compared to $^8$B neutrinos, which have higher energies and are more strongly affected by the modified MSW mechanism.

The electron neutrino survival probability $P_{ee}$ (Eq.(\ref{eqn28})) is calculated assuming adiabatic neutrino propagation.
To justify this assumption, the adiabaticity parameter $\gamma$ must be evaluated and shown to be much greater than one ($\gamma \gg 1$) throughout the neutrino's path.
This parameter is defined as \cite{Strumia}
\begin{eqnarray}\label{eqn29}
\gamma(r^\prime) \equiv \frac{R_\odot \Delta m^2_M(r^\prime)/4E_l(r_B^\prime)}{|d\theta_M /  dr^\prime|} \left[1 + \frac{r^\prime_s}{2r^\prime_B}\right].
\end{eqnarray}
Using the modified mixing angle from Eq.(\ref{eqn26}), the derivative in the denominator becomes $\frac{d\theta_M}{dr^\prime} = \frac{\sin 2\theta }{2 [ 1 + A^2 - 2A \cos 2\theta ] } \frac{dA}{dr^\prime}$, which leads to the general expression for adiabaticity
\begin{eqnarray}\label{eqn30}
\gamma(r^\prime) = \frac{ R_\odot \Delta \kappa_{21}^{\prime 2} \hat{\phi}^4(r_B^\prime) \left[1 + \frac{r_s^\prime}{2 r_B^\prime}\right] \left[1 + A^2 - 2 A\cos2\theta\right]^{\frac{3}{2}}}{2 E_l(r_B^\prime) \sin 2\theta~|\frac{d A}{dr^\prime}|}.
\end{eqnarray}
The most critical condition for adiabaticity occurs at the resonance point.
Here, the expression simplifies since $A_{\text{res.}} \to \cos 2\theta$, yielding
\begin{eqnarray}\label{eqn31}
\gamma_{\text{res.}} = \tilde{\gamma}_\text{{res.}} \left[\frac{\sin^2 2\theta}{2 \pi \cos2\theta}\right],
\end{eqnarray}
where $\tilde{\gamma}_{\text{res.}} = \frac{\pi R_\odot \Delta \kappa_{21}^{\prime 2} \hat{\phi}^4(r_B^\prime)/E_l(r_B^\prime)}{|d\ln A/dr^\prime|_{\text{res.}}} \left[1 + \frac{r_s^\prime}{2 r_B^\prime}\right]$.
The radial derivative $d \ln A/dr^\prime$ is determined by the electron number density profile, $N_e(r^\prime)$.
Adopting the established approximation $N_e(r^\prime) = 245 N_A/\text{cm}^3 \times \exp \left(-10.54~r^\prime \right)$ \cite{Strumia,Bahcall}, where $N_A = 6.022 \times 10^{23} \text{mol}^{-1}$ is Avogadro's constant, the parameter $\tilde{\gamma}_{\text{res.}}$ can be approximated by the expression
\begin{eqnarray}\label{eqn32}
\tilde{\gamma}_{\text{res.}} \approx \frac{\Delta \kappa_{21}^{\prime 2} \hat{\phi}^4(r_B^\prime) / E_l(r_B^\prime)}{10^{-9} \text{eV}^2/\text{MeV}} \left[1 + \frac{r_s^\prime}{2 r_B^\prime}\right].
\end{eqnarray}
For the solar neutrino problem, explained by LMA oscillations with $\Delta m^2_{21} (\text{on Earth}) = \Delta \kappa_{21}^{\prime 2} \hat{\phi}^4(r_B^\prime) \sim 7\times 10^{-5} \text{eV}^2$, the resonance adiabaticity parameter $\gamma_{\text{res.}} > 10^3$ for $^8$B neutrinos with energies around $10~\text{MeV}$.
Therefore, the use of the adiabatic expression in Eq. (\ref{eqn28}) is well justified for the model parameters and environments considered in this study.

\section{Numerical Analysis}\label{sec4}
It is important to note that the survival probability of solar electron-neutrinos, $P_{ee}(E_\nu)$, depends indirectly on the dimensionless parameters $\xi^\prime$ (the scalar-GB invariant coupling) and $r_s^\prime$ (the fractional Schwarzschild radius of sphere).
These parameters might influence the scalar field profile and the curvature of spacetime, which in turn affect neutrino propagation.
On the other hand, $P_{ee}(E_\nu)$ is directly influenced by $\Delta \kappa^{\prime 2}_{21}$, which governs the coupling-squared difference of the interaction between the scalar field and neutrinos.
By analyzing these dependencies, we can gain deeper insights into the alternative gravity theories for neutrino physics and potentially constrain the parameters of the sGB model using solar neutrino experimental data.

The constraints on the parameters of the sGB gravity model are particularly intriguing due to their gravitational interpretations.
One of the key parameters, $\xi = \xi^\prime R_\odot^2$, which determines how the scalar field interacts with the matter components, influences both gravitational and particle physics phenomena.
The choice of a simple quadratic form for the coupling function in Eq. (\ref{eqn2}) ensures that the hairless Schwarzschild and Kerr black hole spacetimes remain valid solutions of the Einstein field equations in the limit of negligible scalar field $\hat{\phi}$ \cite{Doneva,Silva,Herdeiro,Hod}.
This property is essential for maintaining consistency with well-established solutions in general relativity while allowing for new physics in the presence of a non-trivial scalar field.

\paragraph{Statistical Methodology and Likelihood Construction}
The constraints on the model parameters are derived through a least-squares analysis, comparing the theoretical electron-neutrino survival probability $P_{ee}^{\text{th}}(E_\nu)$ to the experimental values obtained from solar neutrino observations.
The dataset includes measurements from the Kamiokande \cite{Kamiokande}, Super-Kamiokande (SK I-IV) \cite{Super-Kamiokande,SK-I,SK-II,SK-III}, Sudbury Neutrino Observatory (SNO Phase I-III, CC and $\nu e$ elastic scattering) \cite{SNO-I,SNO-II,SNO-III}, and Borexino \cite{Borexino-Data,BOREXINO} experiments (listed in the second column of Tables \ref{table1} and \ref{table2}).
For each experiment, the observed survival probability $P_i^{\text{obs}}$ is calculated from the reported neutrino flux $\Phi_i$ normalized by the Standard Solar Model (SSM) [BPS08(GS)] prediction for the $^8$B flux, $\Phi_{\text{SSM}} = (5.94 \pm 0.65) \times 10^6$ cm$^{-2}$ s$^{-1}$ \cite{Pena-Garay,PDG}, such that $P_i^{\text{obs}} = \Phi_i / \Phi_{\text{SSM}}$.
The associated uncertainty $\sigma_i$ is propagated from the experimental and SSM errors.

The compatibility between the model and data is quantified by the $\chi^2$ statistic, defined as
\begin{eqnarray}\label{eqn33}
\chi^2(\zeta) = \sum_{i} \frac{\left(P_i^{\text{obs}} - P_i^{\text{th}}(\zeta)\right)^2}{\sigma_i^2},
\end{eqnarray}
where $\zeta \in  \{\xi^\prime, \Delta\kappa^{\prime 2}_{21}\}$ is considered each model parameter of this study.
The theoretical survival probability $P_i^{\text{th}}$ is computed within the framework of the model for a given value of $\zeta$.

In this construction, all experimental errors are treated as statistical and independent; no correlated systematic uncertainties between different experiments or data releases are considered.
The best-fit values are obtained by minimizing the $\chi^2$-function.
The associated $2\sigma$ and $3\sigma$ confidence intervals are defined by the regions where $\Delta\chi^2(\zeta) = \chi^2(\zeta) - \chi^2(\zeta_{\text{bf}})$ is less than 4.0 and 9.0, respectively.
This analysis employs no external theoretical priors on $\zeta$; the constraints are derived solely from the likelihood of the solar neutrino data.
Furthermore, the results for individual experiments, as well as combined global analyses, are presented in Tables \ref{table1} and \ref{table2}.
The combination of datasets (e.g., Global SK, SK+SNO) is performed by constructing a total $\chi^2$-function that is the sum of the individual $\chi^2$ contributions from each respective experiment before minimization.

As mentioned, these measurements might have profound implications for cosmology and gravity, as they constrain parameters of theories beyond the Standard Model (BSM), such as sGB gravity.
Table \ref{table1} represents a compilation of results of the dimensionless model parameter $\xi^\prime$, a quantity crucial for understanding the scalar field and neutrino interactions.
This table is structured to provide the best-fit values of $\xi^\prime$ alongside its corresponding 2$\sigma$ and 3$\sigma$ ranges.
The best-fit values for $\xi^\prime$ exhibit a high degree of consistency across all experiments, around $\sim 7.8 \times 10^{11}$.
The best-fit values range from approximately $7.632 \times 10^{11}$ (SNO-Phase III (CC)) to $8.240 \times 10^{11}$ (Kamiokande).
The ``Global fit'' across all data yields a best-fit value of $\xi^\prime = 7.834 \times 10^{11}$, which represents the most consistent estimate of the parameter based on the combined data.

The 2$\sigma$ and 3$\sigma$ ranges provide a rigorous quantification of the uncertainties associated with each measurement.
The combined analyses, i.e., Global SK ($0.260$ for 2$\sigma$ and $0.517$ for 3$\sigma$), Global SNO ($0.267$ for 2$\sigma$ and $0.390$ for 3$\sigma$), SK+SNO ($0.306$ for 2$\sigma$ and $0.410$ for 3$\sigma$), and the overall Global fit ($0.294$ for 2$\sigma$ and $0.383$ for 3$\sigma$)—multiplied by a factor $10^{11}$— consistently exhibit the smallest differences in both 2$\sigma$ and 3$\sigma$ ranges.
This signifies that combining data from multiple phases or experiments significantly improves the precision of the $\xi^\prime$-fitting.

\begin{table}[H]
	\begin{center}
		\tiny
		\caption{\footnotesize{Results from solar neutrino experiments regarding $^8$B neutrino flux, the best-fit values and corresponding 2$\sigma$ and 3$\sigma$ confidence intervals have been shown.
				The best-fit values are obtained around $7.8 \times 10^{11}$.
}}
		\label{table1}
		\begin{tabular}{|c|c|c|c|c|}
			\hline
			\hline
			Experiment & $\nu_e$ flux [$10^6/\text{cm}^2 \text{s}$] & $\xi^\prime$ best-fit $[10^{11}]$ &  2$\sigma$ range $[10^{11}]$ & 3$\sigma$ range $[10^{11}]$ \\
			\hline \hline
			Kamiokande \cite{Kamiokande} & $2.80 \pm 0.19$ &$8.240$ & $>7.877$ & $>7.471$ \\ \hline
			SK-I \cite{SK-I} & $2.38 \pm 0.02$ &$8.137$ & $[7.620,8.380]$ & $[7.471,8.588]$ \\
			SK-II \cite{SK-II} & $2.41 \pm 0.05$ &$8.113$ & $[7.605,8.367]$ & $[7.445,8.611]$ \\
			SK-III \cite{SK-III} & $2.32 \pm 0.04$ &$8.035$ & $[7.522,8.265]$ & $[7.394,8.437]$\\
			SK-IV \cite{Super-Kamiokande} & $2.31 \pm 0.02$ &$7.990$ & $[7.489,8.207]$ & $[7.359,8.361]$ \\
			Global SK (SK I-IV) & &$8.087$ & $[7.938,8.198]$ & $[7.739,8.256]$ \\ \hline
			SNO-Phase I (CC) \cite{SNO-I} & $1.76^{+0.06}_{-0.05}$ & $7.745$ & $[7.515,8.046]$ & $[7.411,8.143]$ \\
			SNO-Phase I ($\nu e$) \cite{SNO-I} & $2.39^{+0.24}_{-0.23}$ & $8.085$ & $[7.513,8.449]$ & $[7.399,9.507]$ \\
			SNO-Phase II (CC) \cite{SNO-II} & $1.68 \pm 0.06$ & $7.729$ & $[7.508,8.010]$ & $[7.404,8.107]$ \\
			SNO-Phase II ($\nu e$) \cite{SNO-II} & $2.35 \pm 0.22$ & $8.019$ & $[7.466,8.350]$ & $[7.355,8.861]$ \\
			SNO-Phase III (CC) \cite{SNO-III} & $1.67^{+0.05}_{-0.04}$ & $7.632$ & $[7.419,7.900]$ & $[7.318,7.993]$ \\
			SNO-Phase III ($\nu e$) \cite{SNO-III} & $1.77^{+0.24}_{-0.21}$ & $7.702$ & $[7.465,8.070]$ & $[7.362,8.205]$ \\
			Global SNO &  & $7.747$ & $[7.634,7.901]$ & $[7.587,7.977]$ \\ \hline
			SK+SNO &  & $7.795$ & $[7.676,7.982]$ & $[7.633,8.043]$ \\ \hline
			Borexino  \cite{Borexino-Data,BOREXINO} & $2.57^{+0.17}_{-0.18}$ & $7.770$ & $[7.444,7.925]$ & $[7.334,7.975]$ \\ \hline
			Global fit & & $7.834$ & $[7.698,7.992]$ & $[7.654,8.037]$ \\
			\hline
		\end{tabular}
	\end{center}
\end{table}

The Gauss-Bonnet term, while being a topological invariant in general relativity, naturally interacts with a quintessence scalar field, leading to modifications in gravity at solar system scales \cite{Amendola}.
We can compare the constraints on the model parameter $\xi^\prime$ with findings from various experiments.
These include deviations in planetary motions around the Sun \cite{PlntMotion1,PlntMotion2}, the frequency shift of signals from the Cassini probe \cite{Cassini}, and table-top experiment \cite{tabletop}.
The respective upper bounds are
\[\xi^\prime_{\text{pm}} \lesssim 2.816 \times 10^{16},~~~\xi^\prime_{\text{Cassini}} \lesssim 5.485 \times 10^{12},~~~\xi^\prime_{\text{tt}} \lesssim 3.757 \times 10^{15}.\]
The present model, based on the Gauss-Bonnet gravity driven dark energy, aligns well with all the results except the Kamiokande one, as demonstrated in table \ref{table1}.
Furthermore, we will illustrate that such models typically result in deviations from the standard framework of neutrino oscillations.

Table \ref{table2} includes the results of the neutrino coupling difference $\Delta \kappa^{\prime 2}_{21}$, expressed in units of $[10^{-23} \text{eV}^2]$.
The data originate from fitting of all the aforementioned neutrino oscillation experiments and a Global fit (i.e., combining all the neutrino flux data).
Values of $\Delta \kappa^{\prime 2}_{21}$ range from $0.995 \times 10^{-23} \text{eV}^2$ (SNO-Phase I (CC)) to $1.521 \times 10^{-23} \text{eV}^2$ (Borexino).
The Global fit yields a value of $1.018 \times 10^{-23} \text{eV}^2$, which shows a consistency among the various experimental results.
Combined analyses (Global SK, Global SNO, SK+SNO, Global fit) show tighter constraints compared to individual experiments, demonstrating the power of data combination in reducing uncertainties.

The coupling difference $\Delta \kappa^{\prime 2}_{21}$ is directly related to the mass-squared splitting between the two lightest neutrino mass eigenstates ($\Delta m^{2}_{21}$).
Therefore, accurate determination of this parameter is crucial for understanding the neutrino mass and the properties of neutrino oscillations.

\begin{table}[H]
	\begin{center}
		\tiny
		\caption{\footnotesize{Determinations of $\Delta\kappa^{\prime 2}_{21}$ including best-fit values, 2$\sigma$ and 3$\sigma$ confidence intervals. }}
		\label{table2}
\begin{tabular}{|c|c|c|c|c|}
	\hline
	\hline
	Experiment & $\nu_e$ flux [$10^6/\text{cm}^2 \text{s}$] & $\Delta \kappa^{\prime 2}_{21}$ best-fit $[10^{-23} \text{eV}^2]$ &  2$\sigma$ range $[10^{-23} \text{eV}^2]$ & 3$\sigma$ range $[10^{-23} \text{eV}^2]$ \\
\hline \hline
Kamiokande \cite{Kamiokande} & $2.80 \pm 0.19$ &$1.281$ & $>0.387$ & $>0.089$ \\ \hline
SK-I \cite{SK-I} & $2.38 \pm 0.02$ &$1.019$ & $[0.172,2.181]$ & $[0.098,4.027]$ \\
SK-II \cite{SK-II} & $2.41 \pm 0.05$ &$1.158$ & $[0.199,2.568]$ & $[0.109,5.278]$ \\
SK-III \cite{SK-III} & $2.32 \pm 0.04$ &$1.297$ & $[0.211,2.721]$ & $[0.129,4.585]$\\
SK-IV \cite{Super-Kamiokande} & $2.31 \pm 0.02$ &$1.505$ & $[0.252,3.064]$ & $[0.152,4.944]$ \\
Global SK (SK I-IV) & &$1.159$ & $[0.706,1.660]$ & $[0.354,1.994]$ \\ \hline
SNO-Phase I (CC) \cite{SNO-I} & $1.76^{+0.06}_{-0.05}$ & $0.995$ & $[0.427,2.795]$ & $[0.286,3.841]$ \\
SNO-Phase I ($\nu e$) \cite{SNO-I} & $2.39^{+0.24}_{-0.23}$ & $1.172$ & $[0.157,3.651]$ & $>0.101$ \\
SNO-Phase II (CC) \cite{SNO-II} & $1.68 \pm 0.06$ & $1.009$ & $[0.446,2.664]$ & $[0.300,3.669]$ \\
SNO-Phase II ($\nu e$) \cite{SNO-II} & $2.35 \pm 0.22$ & $1.383$ & $[0.192,3.975]$ & $>0.124$ \\
SNO-Phase III (CC) \cite{SNO-III} & $1.67^{+0.05}_{-0.04}$ & $1.420$ & $[0.632,3.665]$ & $[0.424,5.030]$ \\
SNO-Phase III ($\nu e$) \cite{SNO-III} & $1.77^{+0.24}_{-0.21}$ & $1.213$ & $[0.503,4.305]$ & $[0.337,6.655]$ \\
Global SNO & & $1.034$ & $[0.687,1.777]$ & $[0.578,2.292]$ \\ \hline
SK+SNO & & $1.030$ & $[0.674,1.964]$ & $[0.575,2.406]$ \\ \hline
Borexino \cite{Borexino-Data,BOREXINO} & $2.57^{+0.17}_{-0.18}$ & $1.521$ & $[0.456,2.608]$ & $[0.296,3.086]$ \\ \hline
Global fit & & $1.018$ & $[0.629,1.747]$ & $[0.536,2.025]$ \\
\hline
\end{tabular}
	\end{center}
\end{table}

The bounds on $\xi^\prime$ ($\Delta \kappa^{\prime 2}_{21}$) might have an effect on the present model such that a larger $\xi^\prime$ ($\Delta \kappa^{\prime 2}_{21}$) leads to a case, similar to the neutrino oscillations in vacuum, i.e., $P_{ee} \sim 0.55$, while $P_{ee} \sim 0.33$ for the case with lower $\xi^\prime$ ($\Delta \kappa^{\prime 2}_{21}$).
The constraints on the model parameters $\xi^\prime$ and $\Delta\kappa^{\prime 2}_{21}$ will be used to plot various figures, illustrating the survival probability and the mass function.

\begin{figure}[H]
	\centering
	\includegraphics[scale=0.42]{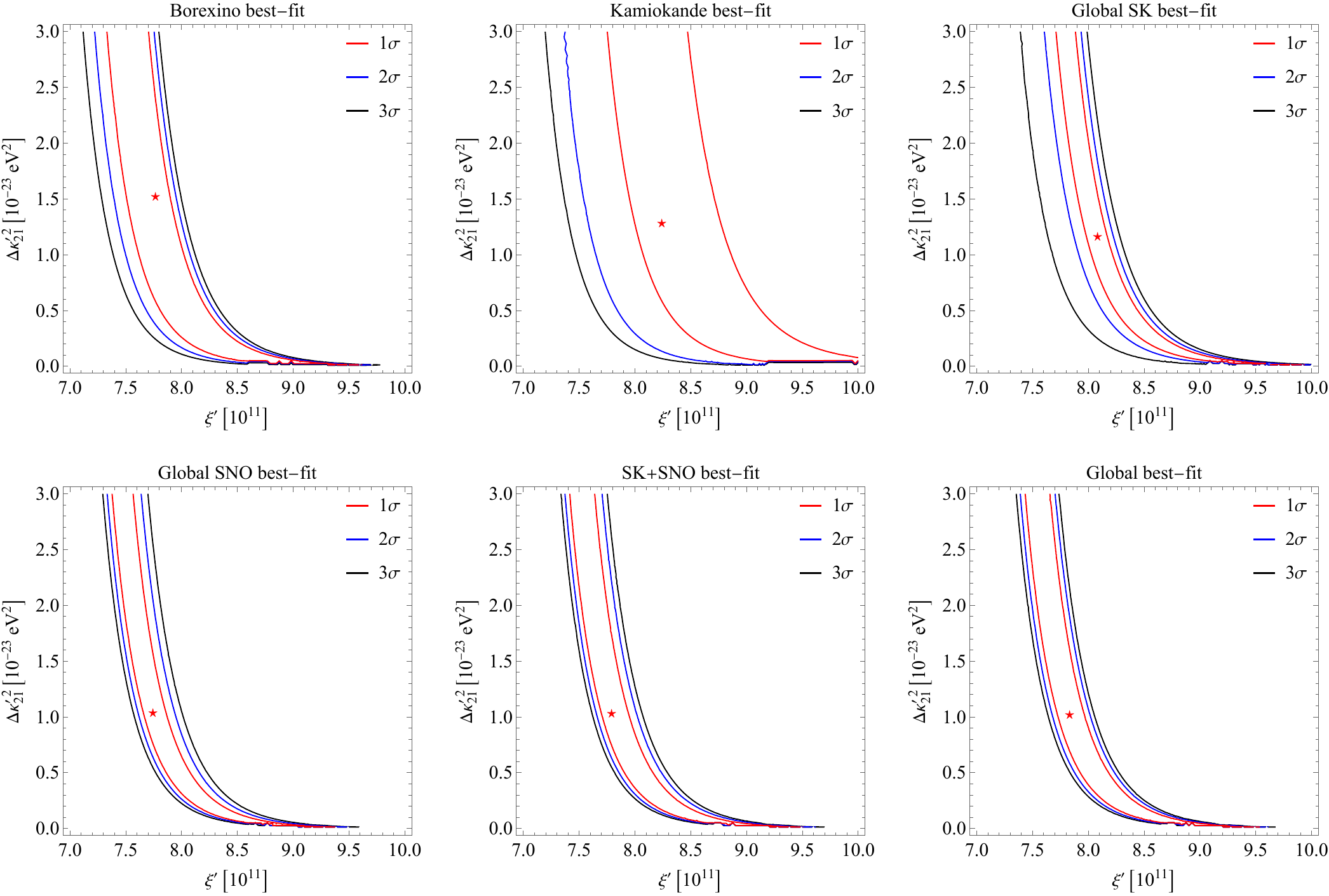}
	\caption{\footnotesize{The analysis of confidence regions (1$\sigma$, 2$\sigma$, and 3$\sigma$) for the model parameters $\xi^\prime$ (dimensionless) and $\Delta \kappa^{\prime 2}_{21}$ (in eV$^2$).
	The best-fit values are shown by asterisks.}}
	\label{fig2}
\end{figure}

The parameter $\xi^\prime$ itself may not be a standard parameter within the simplest 3$\nu$ oscillation framework.
It could be sensitive to extensions of the Standard Model, such as the existence of sterile neutrinos and non-standard interactions \cite{Miranda}.
These constraints on $\xi^\prime$ serve as vital input in the ongoing search for new physics BSM, based on the sGB gravity.
The graphs below show how the constraints on this new physics parameter relate to the on another non-standard oscillation parameter, i.e., $\Delta \kappa^{\prime 2}_{21}$.
The provided plots in Fig.\ref{fig2} depict the confidence regions for the model parameters $\xi^\prime$ and $\Delta \kappa^{\prime 2}_{21}$.

\begin{figure}[H]
	\centering
	\includegraphics[scale=0.5]{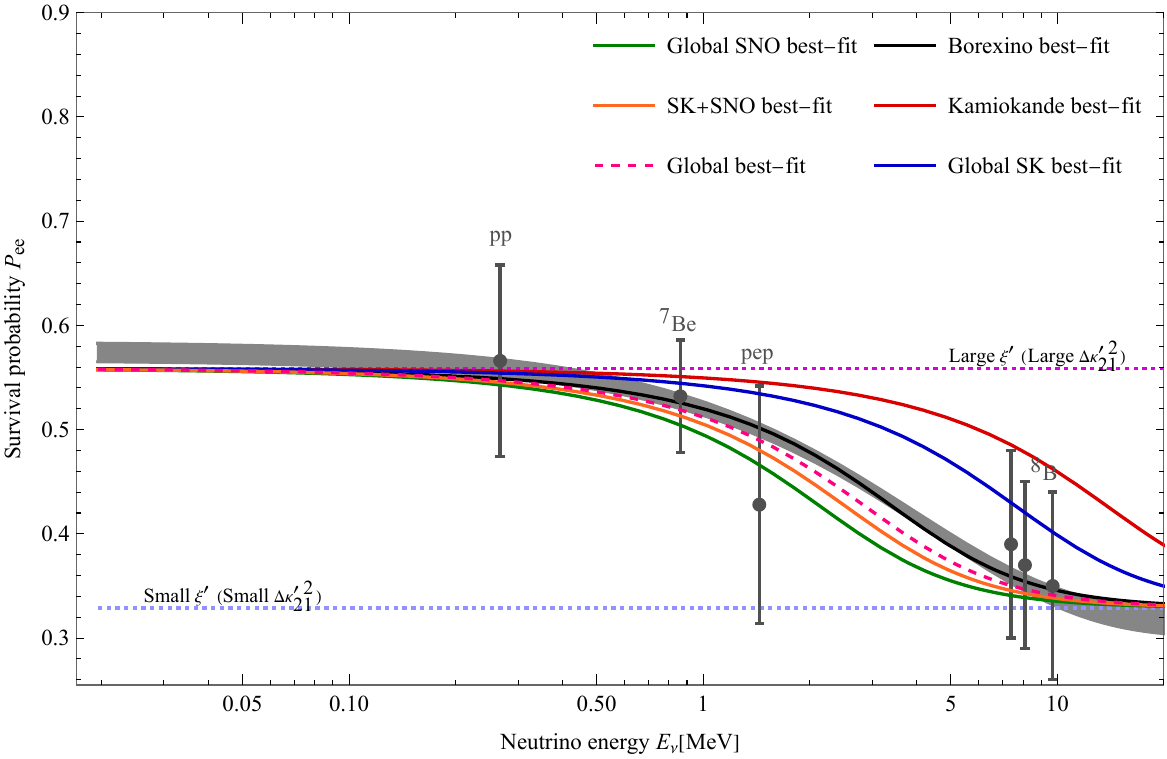}
	\caption{\footnotesize{Survival probability $P_{ee}(E_\nu)$ of solar neutrinos in terms of neutrino energy for LMA-MSW solution, and different curves of LMA-MSW + scalar-neutrino non-standard couplings for six various situations.
	In this figure, the gray band shows the curve from standard matter effects (i.e., the case without scalar field) within $\pm 1\sigma$ C.L., and the horizontal dotted lines show the cases with large/small model parameters $\xi^\prime$ or $\Delta \kappa_{21}^{\prime 2}$.
}}
	\label{fig3}
\end{figure}

From Eq. (\ref{eqn22}), it is expected that non-standard neutrino couplings to a scalar field significantly influence neutrino oscillation probabilities in vacuum. These couplings also alter the survival probability $P_{ee}(E_\nu)$ of solar electron-neutrinos through the LMA-MSW effect as neutrinos traverse solar matter, as illustrated in Fig. \ref{fig3}.
This figure, based on best-fit values from Kamiokande, Global SK, Global SNO, and Borexino spans the energy spectrum of solar neutrinos, including those from $pp$, $pep$, $^7$Be, and $^8$B reactions.
At lower energies (e.g., $pp$ and $^7$Be neutrinos), $P_{ee}$ remains relatively high due to the reduced impact of the MSW effect.
In contrast, at higher energies (e.g., $^8$B neutrinos), $P_{ee}$ decreases significantly as the MSW effect enhances flavor conversion in regions of high electron density within the Sun.
The exploration of physics BSM in solar neutrino flux is most feasible for higher-energy $^8$B neutrinos.
The Borexino detector, designed to detect lowest-energy neutrinos, is susceptible to background noises from radioactive contaminants within its scintillator \cite{Agarwalla} or cosmic ray interactions in the atmosphere \cite{Kumaran}.
These background signals can obscure the true interpretation of results related to physics BSM for lowest-energy neutrinos, i.e., neutrinos from $pp$, $^7$Be, and $pep$ reactions.

The plot serves as a tool to investigate parameters of the present model \cite{Chakraborty,MohseniSadjadi-Yazdani3,MohseniSadjadi-Yazdani4}, obtained above, such as the coupling difference between the scalar field and neutrinos ($\Delta \kappa^{\prime 2}_{21}$) and the non-minimal coupling parameter ($\xi^\prime$).
By comparing best-fit curves with theoretical prediction (the gray band), we can refine the bounds on these parameters and test the validity of the sGB model.
As depicted in Fig. \ref{fig3}, the best-fit curve for Kamiokande (or even Global SK) shows potential discrepancies with the observational data (gray points) \cite{BOREXINO} at neutrino energies near $10$ MeV.
In contrast, the curves derived from the best-fit values of the Global-fit, Borexino, SK+SNO, and Global SNO analyses align well with the measured data points.
The consistency (or discrepancies) between the best-fit curves of different experiments provides valuable insights into the robustness of the LMA solution to the solar neutrino problem.

\begin{figure}[H]
	\centering
	\includegraphics[scale=0.33]{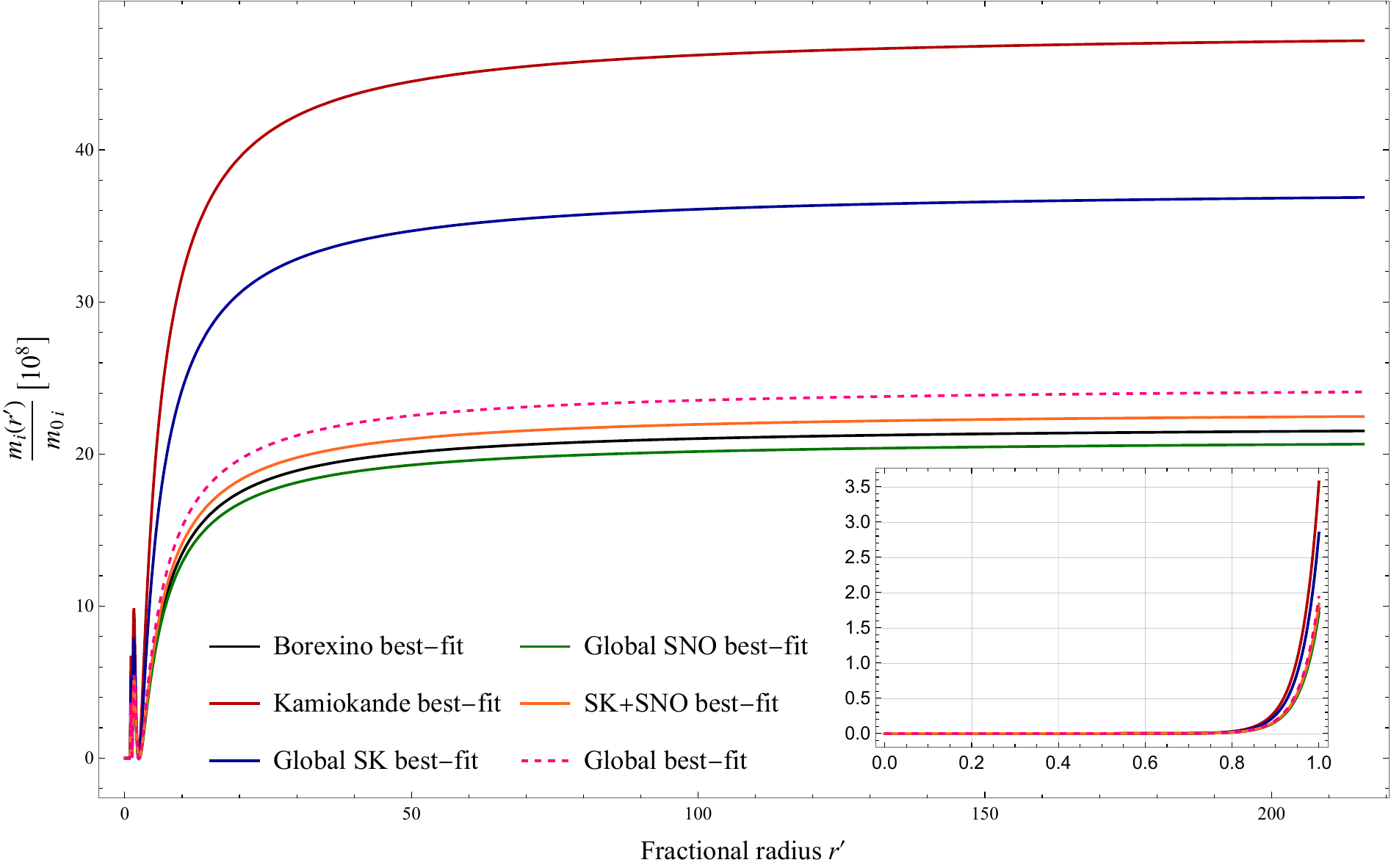}
	\caption{\footnotesize{Figure shows the ratio of mass function of neutrinos in terms of dimensionless radius $r^\prime$, derived from the optimal fits of solar neutrino experiments.
	This figure, in fact, shows the sensitivity of neutrino mass to non-minimal coupling parameter $\xi^\prime$.
	The inset plot illustrates the behavior of the mass function inside the Sun.
}}
	\label{fig4}
\end{figure}

A plot of the ratio of neutrino mass ($m_i(r^\prime)$) to the initial mass of neutrinos ($m_{0i}$), i.e., the fraction $m_i(r^\prime)/m_{0i}$, is illustrated in Fig.\ref{fig4}, ranging from outside the Sun, extending to the Earth, which reflects how neutrino mass changes as neutrinos propagate from the Sun's core to the Earth, influenced by factors such as matter effects and potential new physics caused by non-minimal scalar couplings.

Neutrino mass takes extremely small values as we move closer to the center of the Sun (see the inset plot of Fig. \ref{fig4}); hence, the oscillations of neutrino's flavor inside the object are extremely small.
Outside the Sun, neutrinos propagate through vacuum, where flavor conversion is governed by vacuum oscillations.
The fraction of neutrino mass may exhibit a strongly ascending behavior as a function of distance, reflecting the interference between neutrino mass and scalar field.
Moving further apart, the neutrino mass reaches a limiting value.

The best-fit curves from different experiments show variations due to differences in their sensitivity to non-minimal coupling parameter $\xi^\prime$.
The curve from ``Global SNO best-fit'' may show a less pronounced growth in the fraction of neutrino mass outside the Sun.
The curve from the ``Kamiokande best-fit'', however, represents the most growth in the neutrino mass ratio outside the Sun.

\begin{figure}[H]
	\begin{subfigure}{.5\textwidth}
		\centering
		\includegraphics[scale=0.39]{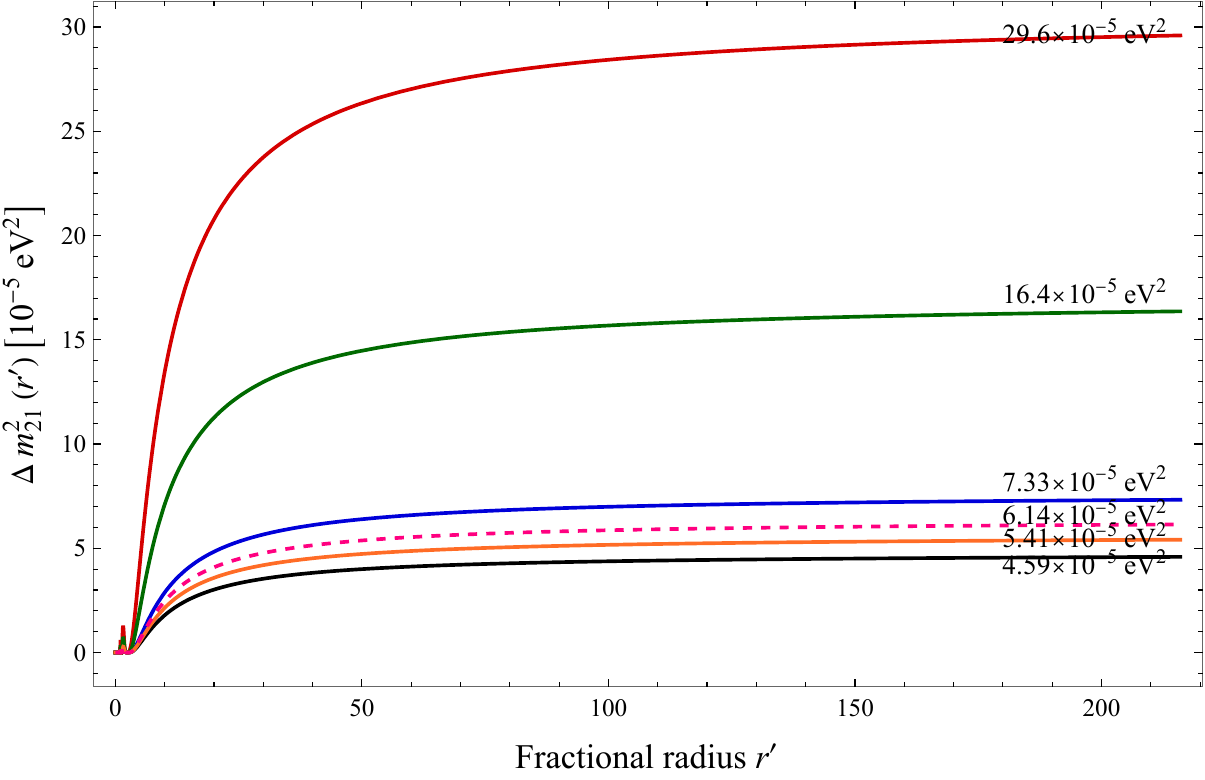}
		\label{fig5-a}
		\caption{\footnotesize{}}
	\end{subfigure}
	\begin{subfigure}{.5\textwidth}
		\centering
		\includegraphics[scale=0.39]{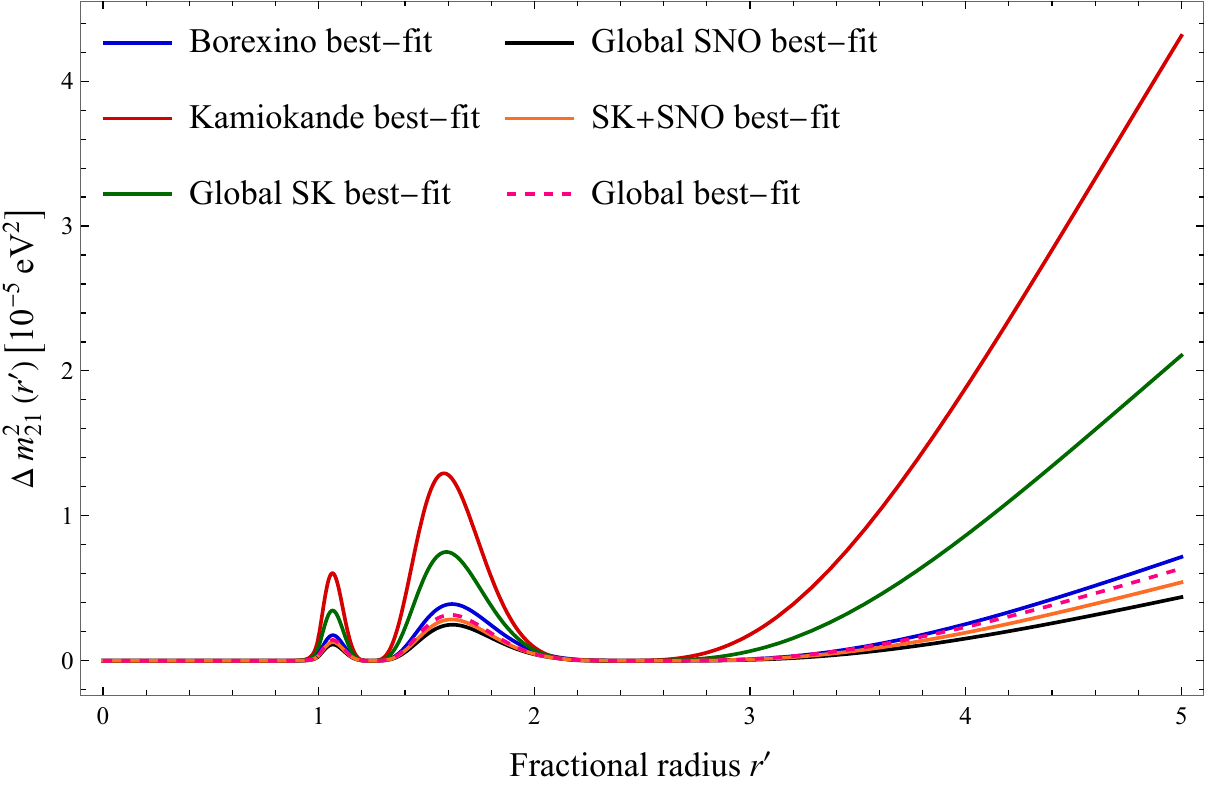}
		\label{fig5-b}
		\caption{\footnotesize{}}
	\end{subfigure}
	\caption{\footnotesize{Variation of the squared mass difference $\Delta m_{21}^2(r^\prime)$ as a function of dimensionless fractional radius ($r^\prime$).
			This would be a comparative analysis from Borexino, SNO, Kamiokande, SK, and Global fits.
			Determined value on each curve indicates the asymptotic mass-squared splitting.}}
	\label{fig5}
\end{figure}

The mass-squared difference $\Delta m^2_{21}$ between neutrino mass eigenstates is a fundamental parameter in neutrino oscillation studies.
The curves observed in $\Delta m^2_{21}$ across fractional radii help refine the understanding of neutrino properties.
Each experiment's best-fit values of model parameters contribute to obtain the neutrino mass differences, see Fig. \ref{fig5}.

The value of mass-squared splitting obtained from the ``Kamiokande best-fit'' data is the largest among the highlighted values, i.e., $29.6 \times 10^{-5} \text{eV}^2$, indicating the most significant mass difference observed in the dataset.
While the corresponding values for Borexino ($7.33 \times 10^{-5} \text{eV}^2$), Global fit ($6.14 \times 10^{-5} \text{eV}^2$), SK+SNO ($5.41 \times 10^{-5} \text{eV}^2$), and Global SNO ($4.59 \times 10^{-5} \text{eV}^2$) best-fits suggest mass differences close to the LMA solution of solar neutrino data \cite{KamLAND,deSalas1,Esteban,deSalas2,Super-Kamiokande,Aharmim1,Bellini,Aharmim2}.

\begin{figure}[H]
	\centering
	\includegraphics[scale=0.43]{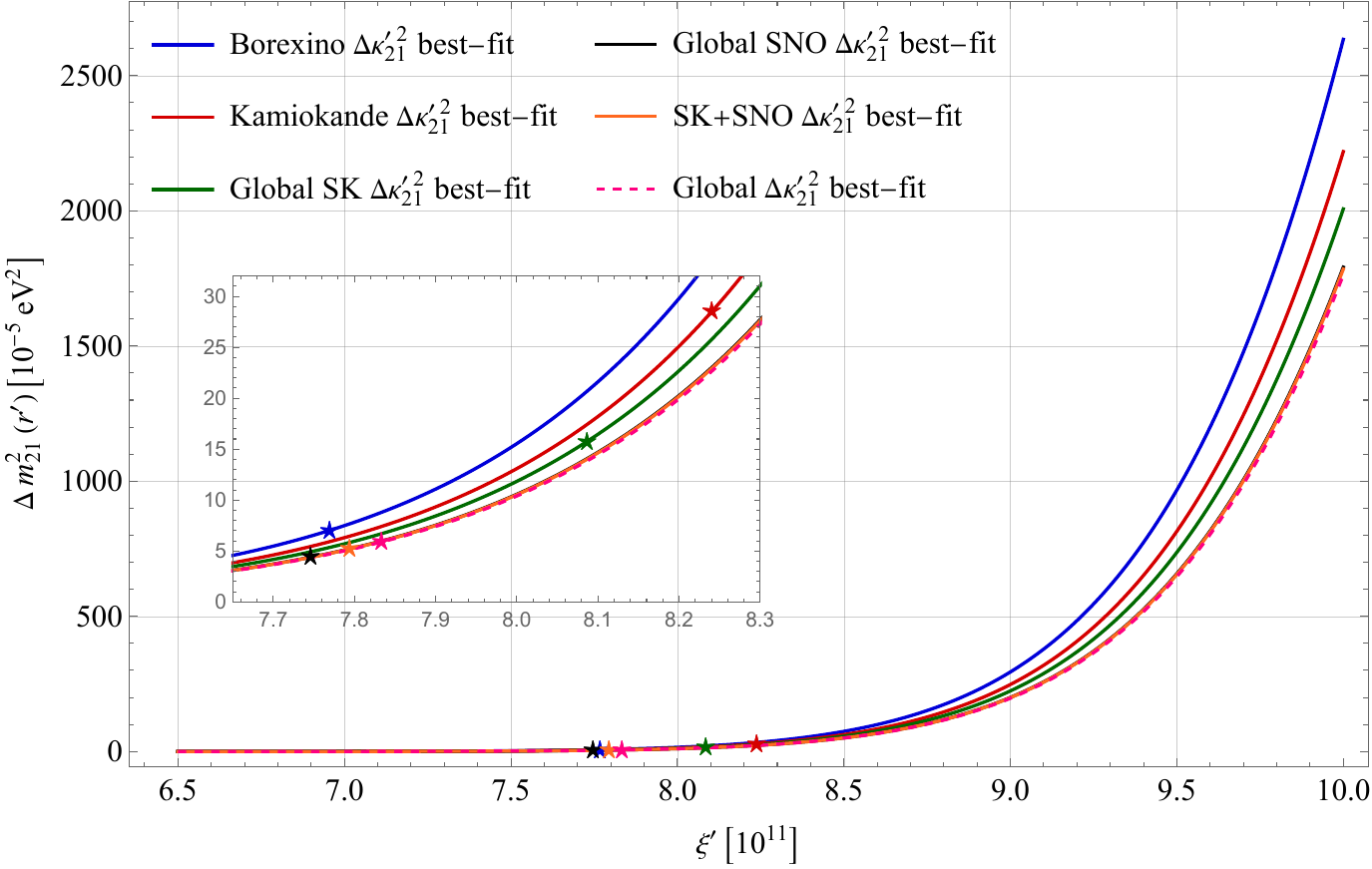}
	\caption{
		\footnotesize{Depiction of $\Delta m_{21}^2$ as it varies with dimensionless model parameter $\xi^\prime [10^{11}]$ as a comparative study featuring the solar neutrino experimetns.
			The asterisks mark the mass-squared splittings $\Delta m_{21}^2$ for the best-fit values of model parameters.}}
	\label{fig6}
\end{figure}

Moreover, figure \ref{fig6} shows variations in $\Delta m^2_{21}$ across different values of $\xi^\prime$ for various best-fits.
The inset plot provides a more detailed view of the variations.
The figure suggests a connection between neutrino mass-squared splittings and the sGB gravity model, mediated by the scalar field and its coupling to the GB invariant.
Also, this could have implications for neutrino physics in strong gravitational fields, such as near neutron stars or black holes.
The asterisks in the plot indicate the corresponding points of the best-fit values ($\xi^\prime$,$\Delta \kappa^{\prime 2}_{21}$) on the mass-squared difference $\Delta m^2_{21}$ curves.
As shown, the best-fit values for ``Kamiokande'' and ``Global SK'' represent the optimal fits, deviating from other data points, might be identified as outlier.

\section{Conclusion}\label{sec5}
We have investigated neutrino flavor conversion within the framework of sGB quadratic theory of gravity.
The quadratic dependence of the neutrino mass term in the action leads to a covariant phase difference, characterized by the emergence of a quartic term of the scalar field in $\Phi_{ij}$.
The environment-dependence of neutrino mass provides a novel mechanism for understanding neutrino oscillations in the presence of gravitational effects.
The hypothesis of environmentally-dependent neutrino mass finds its primary justification in the difference of density in various neutrino propagation environments.

We have derived analytical and numerical solutions for a static, spherically symmetric Schwarzschild spacetime (Section \ref{sec2}).
These solutions demonstrate that the dimensionless scalar field grows from the center of the object to an asymptotic value at large distances, satisfying the boundary conditions.
The phase difference dependence on the quartic order of the scalar field emphasizes the unique role of $\hat{\phi}$ in modifying neutrino
 propagation.
In Section \ref{sec3}, we explored the effects of the scalar-GB invariant and scalar field-neutrino couplings on the MSW effect.
This process can illustrate the interplay between scalar fields and matter effects on neutrino properties,
such as neutrino mass, mass-squared difference and, subsequently, neutrino flavor change.

Using solar neutrino flux measurements from Borexino, Kamiokande, SK, and the SNO, we have constrained the
model parameters $\xi^\prime$ and $\Delta \kappa^{\prime 2}_{21}$, see Section \ref{sec4} and tables therein.
The ``Global fit'', combining data from all experiments, represents the most consistent estimate of the parameters
$\xi^\prime = 7.834 \times 10^{11}$ and $\Delta \kappa^{\prime 2}_{21} = 1.018 \times 10^{-23}$ eV$^{2}$, resulting
in $\Delta m_{21}^2 = 2.87 \times 10^{-23} \text{eV}^2$ inside the Sun and a limiting value $\Delta m_{21}^2 = 6.14 \times 10^{-5} \text{eV}^2$ on the Earth.
These constraints are consistent with the LMA-MSW solution to the solar neutrino problem \cite{KamLAND,deSalas1,Esteban,deSalas2,Super-Kamiokande,Aharmim1,Bellini,Aharmim2} and demonstrate the sGB model's ability to
accommodate observational data without introducing significant deviations from the standard oscillation framework (see figure \ref{fig3}).

The effects of various model parameters, while not yet directly observed, are screened by the precision of current gravitational experiments.
However, future neutrino observations, particularly from next-generation experiments like Hyper-Kamiokande \cite{HK1,HK2}
and DUNE \cite{DUNE}, hold the potential to further test the sGB model indirectly and refine the bounds on its parameters.
Moreover, under the assumption that the spherical object is a black hole, these results may be used to infer properties such as mass and spin, offering a new avenue for probing the nature of black hole candidates \cite{Visinelli}.

In conclusion, this work bridges the gap between neutrino properties and modified gravity, demonstrating the power of
solar neutrino experiments as probes of non-standard interactions.

\section*{Acknowledgement}
This work is based upon research funded by Iran National Science Foundation (INSF) under project No. 4036948.

\end{document}